\documentclass[intlimits,twoside,a4paper]{article}

\usepackage[cp1251]{inputenc}
\usepackage[eqsecnum]{cmpj3}
\usepackage{bm}

\issue{2021}{24}{3}{33605}
\doinumber{10.5488/CMP.24.33605}
\title[Clustering effects on the diffusion of patchy colloids in disordered porous media]%
{ Clustering effects on the diffusion of patchy colloids in disordered porous media\footnote{Dedicated to Yurii Kalyuzhnyi  on the ocassion of his 70th birthday.}}
\author[M.~F. Holovko, M.~Ya. Korvatska]{M.~F. Holovko\orcid{0000-0001-8114-5356},
M.~Ya. Korvatska\orcid{0000-0003-3455-0468}\thanks{Corresponding author: \email{mariamaria301181@gmail.com}.}
}
\address{
	Institute for Condensed Matter Physics of the National Academy of Sciences of Ukraine,
	1 Svientsitskii St., 79011 Lviv, Ukraine
}

\date{Received July 23, 2021, in final form July 28, 2021}

\begin{document}
	
	\maketitle
	
	\begin{abstract}
		Enskog theory is extended for the description of the self-diffusion coefficient of patchy colloidal fluid in disordered porous media. The theory includes the contact values of fluid-fluid and fluid-matrix pair distribution functions that are modified to include the dependence from the so-called probe particle porosity, $\phi$, in order to correctly describe the effects of trapping the fluid particles by a matrix. The proposed expressions for the modified contact values of fluid-fluid and fluid-matrix pair distribution functions include three terms. Namely, a hard sphere contribution obtained by us in the previous work [Holovko M.~F., Korvatska M.~Ya., Condens. Matter Phys., 2020, \textbf{23}, 23605], the depletion contribution connected with the cluster-cluster and cluster-matrix repulsion and the intramolecular correlation inside the cluster. It is shown that the last term leads to a remarkable decrease of the self-diffusion coefficient at a low fluid density. With a decreasing matrix porosity, this effect becomes weaker. For intermediate fluid densities, the depletion contribution leads to an increase of the self-diffusion coefficient in comparison with the hard sphere fluid. For a sufficiently dense fluid, the self-diffusion coefficient strongly decreases due to a hard sphere effect. The influence of the cluster size and the type of clusters as well as of the parameters of porous media is investigated and discussed in detail.
	\end{abstract}

	\section{Introduction}
	It is a great pleasure for us to dedicate this paper to our good friend and colleague Yurii Kalyuzhnyi --- one of the leaders in the area of modern theory of associative fluids. During last years he has successfully applied this theory to the study of the fluids of complex colloidal particles with chemically or physically patterned surfaces~\cite{KalBiaFer15,KalHol14}. These new colloidal systems with strongly anisotropic interparticle interactions are commonly referred as patchy colloids. In particular, in the paper~\cite{KalHol14}, the patchy colloidal system was considered to be confined in the random porous media. The influence of porous media on the phase behaviour and percolation properties of confined patchy colloidal fluids was studied. In the present work, we focus on the investigation of dynamic properties of patchy colloidal fluids in disordered porous media.
	During last three decades, starting from the pioneering work of Madden and Glandt~\cite{MadGlandt88}, much theoretical efforts have been devoted to the study of fluids confined in disordered porous media. In this work~\cite{MadGlandt88}, a disordered porous medium is considered as a matrix of quenched configuration of randomly distributed obstacles. Using the replica Ornstein-Zernike (ROZ) integral equation theory~\cite{GivenStell92}, the statistical mechanical approach for liquid state was extended to describe different models of fluids confined in disordered porous media~\cite{Ros99,Pizio00}, including the chemical reacting fluids~\cite{Trokh96,Trokh97}. However, in contrast to the bulk fluids, no analytical results have been obtained within ROZ approach even for a simple model like hard sphere fluid in a hard-sphere matrix.
	In order to obtain the analytical results, Holovko and Dong~\cite{HolDong09} have recently proposed to extend the classical scaled particle theory (SPT)~\cite{ReissFrisLeb59} for the description of thermodynamic properties of hard sphere fluids in disordered porous media. During the last decade, the SPT approach for hard sphere fluids in disordered porous media was essentially improved~\cite{ChenDong10,HolShmot10,PatHol11,HolPat12,HolPatDong12,HolPatDong17}. The approach proposed in~\cite{HolDong09}, referred to as SPT1, contains a subtle inconsistency manifested when the size of matrix particles is considerably larger than the size of fluid species. This inconsistency was eliminated in a new version labled as SPT2~\cite{PatHol11}. As a result, the first rather accurate analytical expressions were obtained for the chemical potential and pressure of a hard sphere fluid in a hard sphere matrix. These expressions include three parameters describing the porosity of the matrix. The first one is related to the bare geometry of the matrix. It is the so-called geometrical porosity, $\phi_{0}$, that characterizes the free volume not occupied by the matrix particles. The second parameter, $\phi$, is defined by the chemical potential of a fluid in the limit of infinite dilution. It is the so-called probe particle porosity, which means the probability to find the fluid particle in an empty matrix. The third parameter, $\phi^{*}$, is determined by the maximum value of the fluid packing fraction of a hard sphere fluid in a porous medium. It characterizes the maximum adsorption capacity of a matrix for a given type of a fluid. The SPT2 approach was generalized for fluids of anisotropic particles~\cite{HolShmot14,HolShmot18}, for a hard sphere mixture~\cite{ChenZhao16}, for a mixture of hard spheres and anisotropic particles~\cite{HvozdPat18} in disordered porous media. The obtained results were also used as a reference system to take into account the attractive~\cite{HolPat15,HolShmot20}, the associative~\cite{KalHol14} and Coulomb~\cite{HolPatPat16,HolPatPat17,HolovPatPat17,PatPatHolov18,PatPatHol18} interactions. The SPT2 approach is useful not only for the description of thermodynamic properties, but it can be also applied for the description of the static structure of hard sphere fluids in disordered porous media. In particular, a simple and rather accurate expression for the contact value of the fluid-fluid pair distribution function was obtained, which was then used for the description of the phase behaviour and percolation properties of patchy colloidal fluids in disordered porous media~\cite{KalHol14}.
	The structural information provided by the SPT2 approach can be used as an input into the Enskog theory for the description of dynamic properties of hard sphere fluids in a porous media. For this purpose in~\cite{HolovKor20}, the fluid confined in a disordered matrix was considered as a mixture of two components. One of them is quenched in space and is treated as particles with infinite mass. In such an approach, the contact values of the fluid-fluid and fluid-matrix pair distribution functions should be introduced as input to the theory. Using the SPT2 approach, a simple analytical expression for the contact values of both distribution functions was obtained in~\cite{HolovKor20}. However, the obtained expressions describe the influence of the porous media only due to geometric porosity, $\phi_{0}$, and do not include the effects of other types of porosity that are important for the description of thermodynamic properties. It was shown that the application of such contact values neglects  trapping of fluid particles in a matrix and at least the probe particle porosity, $\phi$, should be included in the Enskog theory for a correct description of the matrix influence. In~\cite{HolovKor20}, the Enskog theory was extended by changing the contact values of the fluid-matrix and fluid-fluid pair distribution functions by their counterparts which include  geometrical porosity, $\phi_{0}$, and the probe particle porosity, $\phi$. It was shown that such an improvement of the Enskog theory corresponds to the SPT2b1 approximation for the description of thermodynamic properties. This level of theory predicts correct trends for the influence of porous media on the self-diffusion coefficient of a hard sphere fluid in disordered porous media in a good agreement with computer simulations~\cite{ChangJad04}.
	In the present paper, we generalize the results obtained in~\cite{HolovKor20} within the Enskog theory of dynamic properties of the patchy colloidal fluid in disordered porous media. We model a fluid of associating hard spheres with association due to highly directional, attractive interaction leading to the formation of colloidal clusters. Dependent on the peculiarity of inter-colloidal attraction, clusters of different shape can be formed (e. g., dimers, linear or branched polymer chains, network-like aggregates, star polymers, etc.). We focus on the influence of the clustering effects on the self-diffusion coefficient of patchy colloids. Similarly to~\cite{HolovKor20}, with this aim we need the contact values of the fluid-fluid and fluid-matrix pair distribution functions. In the present study, we restrict ourselves to linear chains and four-fold coordinated network-like clusters. To describe the fluids with such a type of clusters (in the bulk case and near the solid surface)~\cite{KalStellHol95,HolovVakDud95,HolovVak96,VakHol97,VakDudaHolovko97,VoronovLuzMinSid97,VakH97,VakH03}, simple analytical expressions for the contact values of the fluid-fluid pair distribution function and for the wall-fluid density profile in the associative Percus-Yevick approximation have been obtained. We note that the expression for the profile was obtained in the framework of the associative version of Henderson-Abraham-Barker approach~\cite{HolV95}. By simple modification of these expressions, we present the contact values for the fluid-fluid and fluid-matrix pair distribution functions which are used to study the influence of the clustering effects on the self-diffusion coefficient of the patchy colloids in disordered porous media.
	The paper is organized as follows, a brief review of our previous results~\cite{HolovKor20} for the diffusion of the hard-sphere fluid in random porous media is presented in section~2. In addition, we present a generalisation of these results for the associating hard sphere fluid in random porous media. In section~3, some numerical analyses of the influence of clustering and of porous medium on the self-diffusion coefficient are presented. Finally, we draw some conclusions in the last section.
	
	\section{The Enskog theory description of the diffusion of the patchy colloids in disordered porous media}

	In this work we generalize the results~\cite{HolovKor20} for the diffusion of a hard sphere fluid in disordered porous media for the patchy colloidal fluid. Similarly to~\cite{HolovKor20}, the patchy colloidal fluid in disordered porous media is modelled by an equilibrium mixture of associative hard spheres and hard spheres in which the latter component is quenched in space and is considered as particles with infinite mass. As a result, after the application of the Enskog theory, the following expression for the self-diffusion coefficient of the patchy colloidal fluid was obtained~\cite{HolovKor20},
	\begin{equation}
	D_{1}/D_{1}^{0}=\frac{\sqrt{2\piup}}{32} \left[ \frac{1}{\sqrt{2}}\eta_{1}g_{11}(\sigma_{11})+\frac14\tau(\tau+1)^{2}\eta_{0}g_{10}(\sigma_{10})\right]^{-1}
	,\label{HolKor2.1}
	\end{equation}
	where $D_{1}^{0}= \left( \frac{kT\sigma_{11}^{2}}{m_{1}}\right) ^{1/2}$, $T$ is the temperature, $k$ is the Boltzmann constant, $m_1$ is the mass of fluid particle. We use a conventional notation: the index ``1'' denotes the fluid component and the index ``0'' denotes the matrix particles~\cite{GivenStell92,Ros99,Pizio00}. Thus,
	$\eta_{1}=1/6\piup\rho_{1}\sigma_{11}^{3}$ and $\eta_{0}=1/6\piup\rho_{0}\sigma_{00}^{3}$ are the packing fraction of the fluid and matrix species, respectively ($\rho_{1}$ and $\rho_{0}$ are the number densities of fluid and matrix particles, $\sigma_{11}$ and $\sigma_{00}$ are the diameters of fluid and matrix particles). Moreover, $\sigma_{10}=1/2\left( \sigma_{00}+\sigma_{11}\right)$; $\tau=\sigma_{11}/\sigma_{00}$; $g_{11}(\sigma_{11})$ and $g_{10}(\sigma_{10})$ are the contact values of the fluid-fluid and fluid-matrix pair distribution functions.
	To proceed, the expressions for the contact values $g_{11}(\sigma_{11})$ $g_{10}(\sigma_{10})$ are needed. In~\cite{HolovKor20} using the so-called scaled particle theory within SPT2 approach, the following expressions for the $g_{11}(\sigma_{11})$ and $g_{10}(\sigma_{10})$ for a hard sphere fluid in random porous media were obtained,
	\begin{equation}
	 g_{11}(\sigma_{11})=\frac{1}{\phi_{0}-\eta_{1}}+\frac32\frac{\eta_{1}+\eta_{0}\tau}{(\phi_{0}-\eta_{1})^{2}}+\frac12\frac{(\eta_{1}+\tau\eta_{0})^{2}}{(\phi_{0}-\eta_{1})^{3}}
	,\label{HolKor2.2}
	\end{equation}
	\begin{equation}
	 g_{10}(\sigma_{10})=\frac{1}{\phi_{0}-\eta_{1}}+\frac3{(1+\tau)}\frac{\eta_{1}+\eta_{0}\tau}{(\phi_{0}-\eta_{1})^{2}}+\frac2{(1+\tau)^{2}}\frac{(\eta_{1}+\tau\eta_{0})^{2}}{(\phi_{0}-\eta_{1})^{3}}
	.\label{HolKor2.3}
	\end{equation}
	As we can see, the obtained expressions~(\ref{HolKor2.2}) and~(\ref{HolKor2.3}) contain only the geometrical porosity,
	\begin{equation}
	\phi_{0}=1-\eta_{0}
	,\label{HolKor2.4}
	\end{equation}
	and do not include the dependence on other porosity parameters, such as the so-called probe particle porosity, $\phi$. The latter is important for the description of thermodynamic properties of a fluid in disordered porous media. It was shown in~\cite{HolovKor20}, that the application of the contact values given by (\ref{HolKor2.2}) and~(\ref{HolKor2.3}) neglects the effects of trapping of fluid particles by the matrix. At least the so-called probe particle porosity,
	\begin{equation}
	\phi=\phi_{0}\exp\left[ -\frac{\eta_{0}\tau}{1-\eta_{0}}\left( 3(1+\tau)+\frac92\tau\frac{\eta_{0}}{1-\eta_{0}}+\frac{1+\eta_{0}+\eta_{0}^{2}}{(1-\eta_{0})^{2}}\tau^{2}\right) \right]
	,\label{HolKor2.5}
	\end{equation}
	should be included in the Enskog theory for a correct description of the matrix effect. The contact values of the fluid-fluid and fluid-matrix pair distribution functions have proven to be in a very good agreement with computer simulations data~\cite{KalHol14,HolovKor20,ChangJad04}. However, the Enskog theory based only on static correlations between particles is not completely appropriate to describe the effect of disordered matrix and requires modification.
	In~\cite{HolovKor20}, the Enskog theory was extended by modifying the contact values of the fluid-matrix and fluid-fluid distribution functions by their counterparts that include the dependence not only on geometrical porosity, ${\phi_{0}}$, but also on the so-called probe particle porosity, $\phi$. In such a procedure, in the limit $\phi\to\phi_{0}$, these fictitious  contact values coincide with the corresponding ones from the SPT2 approach. It should be noted that the correction of the fluid-matrix contact value is more important than the correction of fluid-fluid contact value. The first of them is very important for the description of the diffusion coefficient for all fluid densities. On the other hand, the second correction is important to describe the density dependence of the self-diffusion coefficient. Having this in mind, in~\cite{HolovKor20}, only the fluid-matrix contact value was modified by the expression,

	\begin{equation}
	g_{10}(\sigma_{10})=\frac{\phi_{0}}{\phi}\left[\frac{1}{1-\eta_{1}/\phi}+\frac3{(1+\tau)}\frac{{\eta_1}/{\phi_{ 0}}}{\left( 1-\eta_{1}/\phi_{0}\right) ^{2}}
	+\frac2{(1+\tau)^{2}}\frac{({\eta_{1}}/{\phi_{0}})^{2}}{\left( 1-\eta_{1}/\phi_{0}\right) ^{3}}\right]
	.\label{HolKor2.6}
	\end{equation}
	Since the first term in~(\ref{HolKor2.6}) leads to the divergence at $\eta_1=\phi$, this term is corrected in close similarity to thermodynamic considerations,
	\begin{equation}
	\frac{1}{1-\eta_{1}/\phi}\to
	\frac{1}{1-\eta_{1}/\phi_{0}}+ \frac{\eta_{1}(\phi_{0}-\phi)}{\phi_{0}\phi(1-\eta_{1}/\phi_{0})^{2}}
	,\label{HolKor2.7}
	\end{equation}
	As a result,
	\begin{equation}
	g_{10}(\sigma_{10})=\frac{\phi_{0}}{\phi}\left[\frac{1}{1-\eta_{1}/\phi_{0}}+ \frac{\eta_{1}(\phi_{0}-\phi)}{\phi_{0}\phi(1-\eta_{1}/\phi_{0})^{2}}+\frac3{(1+\tau)}\frac{{\eta_1}/{\phi_{0}}}{\left( 1-\eta_{1}/\phi_{0}\right) ^{2}}
	+\frac2{(1+\tau)^{2}}\frac{({\eta_{1}}/{\phi_{0}})^{2}}{\left( 1-\eta_{1}/\phi_{0}\right) ^{3}}\right]
	.\label{HolKor2.8}
	\end{equation}
	Still, for $g_{11}(\sigma_{11})$ we use the expression~(\ref{HolKor2.2}).
	 According to~\cite{HolovKor20}, the expression~(\ref{HolKor2.1}) for the self-diffusion of a hard sphere fluid in porous media together with~(\ref{HolKor2.2}) for $g_{11}(\sigma_{11})$ and  with~(\ref{HolKor2.8}) for $g_{10}(\sigma_{10})$ correctly reproduce the effect of porous media on the diffusion coefficient.
	Now, we would like to take into account the clustering effects for the contact values $g_{11}(\sigma_{11})$ and $g_{10}(\sigma_{10})$. As the starting point with this aim, we consider the analytical expressions for the contact values of the fluid-fluid distribution function and fluid-wall density profile for associating fluids obtained in associative Percus-Yevick approximation from~\cite{VakH03},
	\begin{equation}
	g_{11}(\sigma_{11})=\frac{1-\eta_{1}/2}{(1-\eta_{1})^{3}}-\frac{S_{1}-1}{S_{1}}\frac{1}{1-\eta_{1}}+\frac{C}{\eta_{1}} \frac{(S_{1}-1)^{2}}{S_{1}^{2}}
	,\label{HolKor2.9}
	\end{equation}
	\begin{equation}
	 \rho_{1}(\sigma_{11}/2)=\rho_{1}g_{10}(\sigma_{10})={\rho_{1}}\left[\frac{1+2\eta_{1}}{(1-\eta_{1})^{2}}-\frac{S_{1}-1}{S_{1}}\frac{1}{1-\eta_{1}}\right]
	,\label{HolKor2.10}
	\end{equation}
	where $S_{1}$ is the mean cluster size for associating fluid and can be considered as the mean number of a hard sphere monomers that belong to a cluster.
	In both expressions,~(\ref{HolKor2.9}) and~(\ref{HolKor2.10}), the first term corresponds to a hard sphere packing effects which dominate with an increasing $\eta_{1}$. While generalizing~(\ref{HolKor2.9}) and~(\ref{HolKor2.10}) for the contact values $g_{11}(\sigma_{11})$ and $g_{10}(\sigma_{10})$ for associating hard sphere fluid in random porous media, we should change their first terms into~(\ref{HolKor2.2}) and (\ref{HolKor2.8}), correspondingly. The second term in each of the equations~(\ref{HolKor2.9}) and~(\ref{HolKor2.10}) describe the depletion effects due to cluster-wall (matrix) or cluster-cluster repulsion with an increase of $S_1$. While generalizing~(\ref{HolKor2.9}) and~(\ref{HolKor2.10}), we should apply the trick similarly to~(\ref{HolKor2.6}),
	\begin{equation}
	-\frac{S_{1}-1}{S_{1}}\frac{1}{1-\eta_{1}}\to
	-\frac{S_{1}-1}{S_{1}}\frac{\phi_{0}}{\phi}\frac1{1-\eta_{1}/\phi}
	,\label{HolKor2.11}
	\end{equation}
	with the dominator $1/(1-\eta_{1}/\phi$) which should be represented in the form (\ref{HolKor2.7}).
	The last term in (\ref{HolKor2.9}) for $g_{11}(\sigma_{11})$, is responsible for the intramolecular correlation inside a cluster and does not modify after mapping $g_{11}(\sigma_{11})$ for associating hard sphere fluid in random porous media. At fixed values of parameters, $\eta_1$ and $S_1$, the intensity of the intramolecular correlation is given by the coefficient, $C$,~\cite{VakDudaHolovko97}. For chains $C=1/24$ and for networks $C=1/16$~\cite{VakDudaHolovko97,VoronovLuzMinSid97,VakH03}. Therefore, in the considered approximation, the network-like clusters differ from linear chains only by the strength of the intramolecular correlation which dominates at low densities, i.e., when the cluster-cluster contacts are rather rare and monomers correlate mostly through the cluster connectivity. Such a mechanism, as it was shown in~\cite{VoronovLuzMinSid97}, is responsible for the remarkable increase of the surface coverage at low polymer concentrations. It is also important for the explanation of specific features of diffusion behaviour of associated fluids at low concentrations. This issue is discussed in the next section.
	Finally, for the contact values of fluid-fluid and fluid-matrix, we  apply the following expressions,
	\begin{equation}
	 g_{11}(\sigma_{11})=\frac{1}{\phi_{0}-\eta_{1}}+\frac32\frac{\eta_{1}+\eta_{0}\tau}{(\phi_{0}-\eta_{1})^{2}}+\frac12\frac{(\eta_{1}+\tau\eta_{0})^{2}}{(\phi_{0}-\eta_{1})^{3}}-{\frac{S_{1}-1}{S_{1}}\frac{1}{\phi_{0}-\eta_{1}}+\frac{C}{\eta_{1}} \frac{(S_{1}-1)^{2}}{S_{1}^{2}}}
	,\label{HolKor2.12}
	\end{equation}
	\begin{eqnarray}
	\nonumber
	g_{10}(\sigma_{10})&=&\frac{\phi_{0}}{\phi}\left[\frac{1}{1-\eta_{1}/\phi_{0}}+ \frac{\eta_{1}(\phi_{0}-\phi)}{\phi_{0}\phi(1-\eta_{1}/\phi_{0})^{2}}+\frac3{(1+\tau)}\frac{{\eta_1}/{\phi_{0}}}{\left( 1-\eta_{1}/\phi_{0}\right) ^{2}}
	+\frac2{(1+\tau)^{2}}\frac{({\eta_{1}}/{\phi_{0}})^{2}}{\left( 1-\eta_{1}/\phi_{0}\right) ^{3}}\right.\nonumber\\
	&-&\left.{\frac{S_{1}-1}{S_{1}}}{\left(\frac{1}{1-\eta_{1}/\phi_{0}}+ \frac{\eta_{1}(\phi_{0}-\phi)}{\phi_{0}\phi(1-\eta_{1}/\phi_{0})^{2}}\right)}\right]
	.\label{HolKor2.13}
	\end{eqnarray}

	However, we should note that the mean cluster size, $S_1$, in the considered approach is chosen as an external parameter. This means that we do not consider the clustering process itself and assume that the cluster composition remains fixed. More generally, $S_1$, should be considered as a function of density and temperature of the fluid and of the parameters of porous media such as $\eta_{0}$ and $\tau$. Then, the expressions~(\ref{HolKor2.12})-(\ref{HolKor2.13}) remain applicable with $S=S(\eta_{1};T;\eta_{0};\tau)$. In this case, the difference between chains and network clusters would be more pronounced. In particular, for networks, $S_1$, could diverge signalling the percolation threshold~\cite{VakDudaHolovko97,KalHol14,VakDudHolov97,KalJacovDochHolCum11}.

	\section{Results and discussion}
	Based on the expression~(\ref{HolKor2.1}) for the patchy colloidal fluid in disordered porous media and using~(\ref{HolKor2.12}) and~(\ref{HolKor2.13}), in this section we investigate the influence of clustering and porous media on the self-diffusion coefficient. We note that the diffusion of the fluid is defined by the denominator of the expression~(\ref{HolKor2.1}) which can be written in the form,
	
	\begin{eqnarray}
	\nonumber
	D_{n}&=&\frac{1}{\sqrt{2}}\eta_{1}g_{11}(\sigma_{11})+\frac14\tau(\tau+1)^{2}\eta_{0}g_{10}(\sigma_{10})\\
	 &=&\frac{1}{\sqrt{2}}\left[C(1-X)^{2}+\frac{X\eta_{1}}{\phi_{0}-\eta_{1}}+\frac32{\eta_{1}}\frac{\eta_{1}+\eta_{0}\tau}{(\phi_{0}-\eta_{1})^{2}}+\frac12{\eta_{1}}\frac{(\eta_{1}+\tau\eta_{0})^{2}}{(\phi_{0}-\eta_{1})^{3}}\right]\nonumber \\
	 &+&\frac14{\tau}{(\tau+1)^2}\eta_{0}\frac{\phi_{0}}{\phi}{\left[X\frac{1}{1-\eta_{1}/\phi_{0}}(1+\frac{\eta_{1}(\phi_{0}-\phi)}{\phi_{0}\phi(1-\eta_{1}/\phi_{0})^{2}})+\frac3{(1+\tau)}\frac{{\eta_1}/{\phi_{0}}}{\left(1-\eta_{1}/\phi_{0}\right)^{2}}\right.}\nonumber\\
	&+&\left.\frac2{(1+\tau)^{2}}\frac{({\eta_{1}}/{\phi_{0}})^{2}}{\left( 1-\eta_{1}/\phi_{0}\right)^{3}}\right]
	,\label{HolKor3.1}
	\end{eqnarray}
	where $X=1/{S_1}$ is the fraction of the nonbonded patches.
	We begin from the bulk case. With this aim, we put $\eta_{0}=0$ in (\ref{HolKor2.1}) and the diffusion coefficient then is defined only by the contact value,
	\begin{equation}
	g_{11}(\sigma_{11})=\frac{1-\frac12\eta_{1}}{(1-\eta_{1})^3}-\frac{1-X}{1-\eta_{1}}+\frac{C}{\eta_{1}}({1-X})^{2}
	.\label{HolKor3.2}
	\end{equation}

\begin{figure}[h]
		\centerline{
			\includegraphics [height=0.42\textwidth]{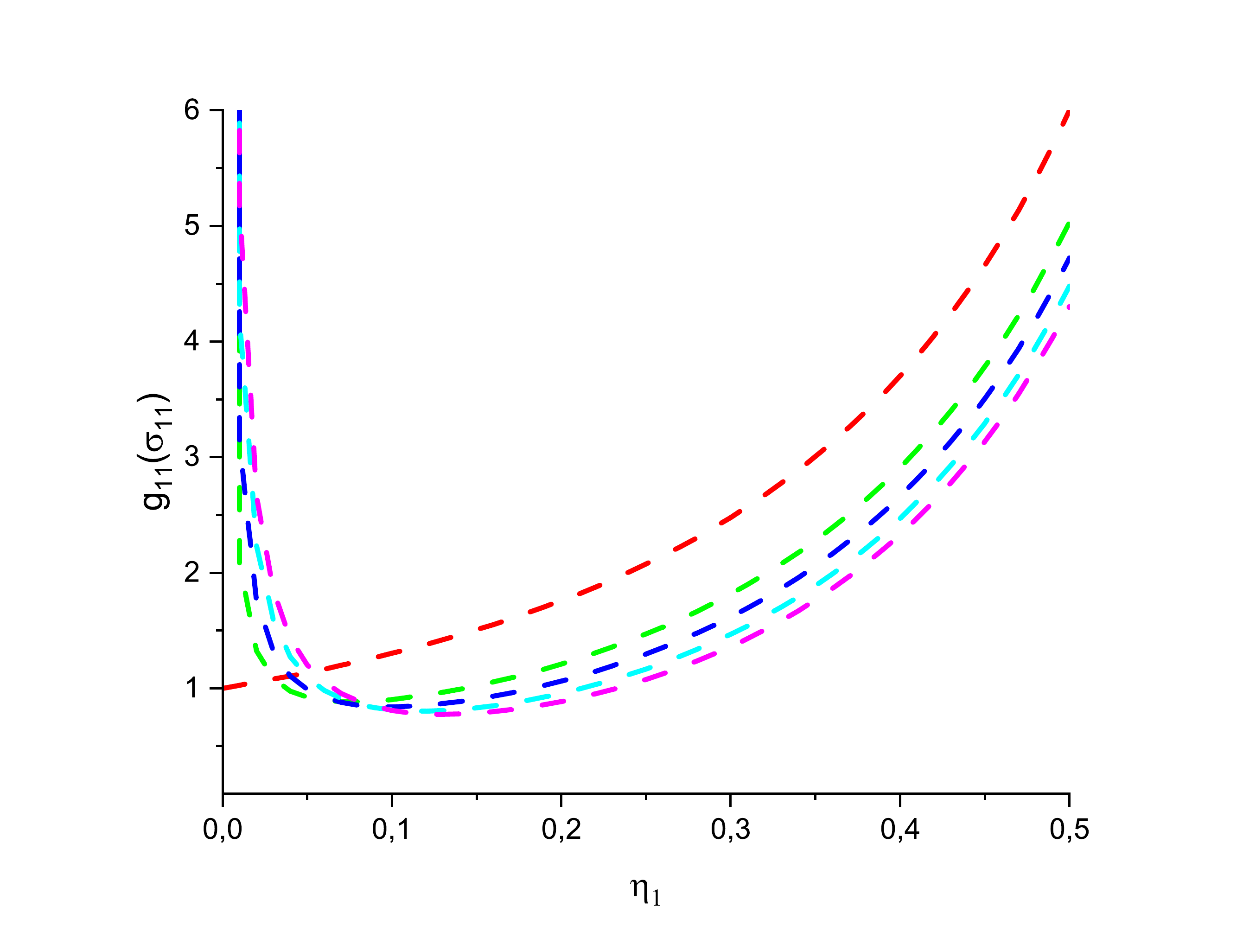}
			\includegraphics [height=0.42\textwidth]{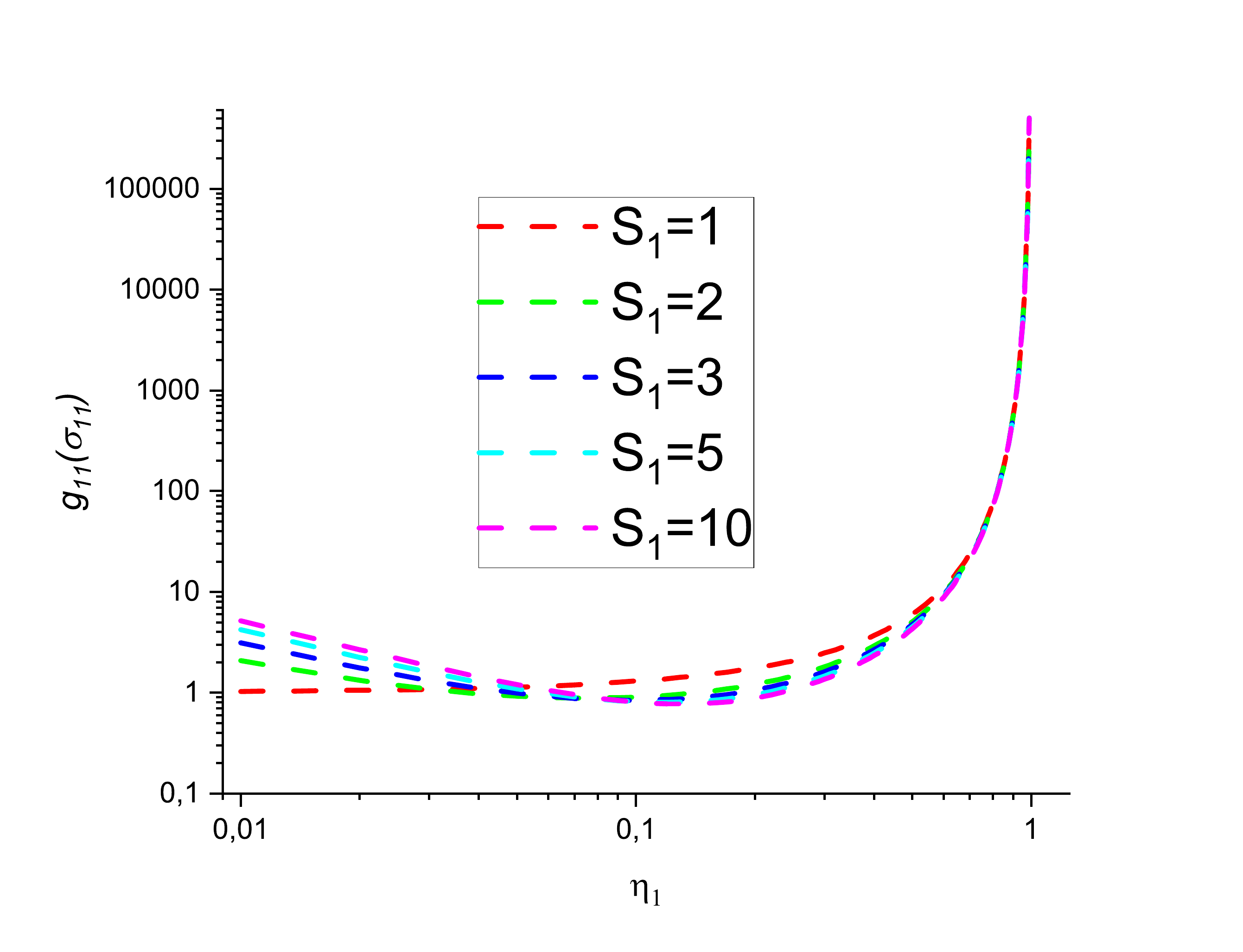}
		}
		\caption{(Colour online) Panel a: The dependence of the contact value of fluid-fluid distribution function $g_{11}(\sigma_{11})$ for network-like fluid on the fluid packing fraction $\eta_1$ for the bulk fluid at a fixed cluster size $S_1$. Panel b: the same as in the panel a, but in the logarithmic scale for better visualization. The nomenclature of lines is the same in both panels.}
		\label{FIG1}
	\end{figure}

	The dependence of the contact value $g_{11}(\sigma_{11})$ on the packing fraction $\eta_{1}$ for the network-like liquid in the bulk case is shown in figure~\ref{FIG1}. For the convenience of a reader this dependence is given in logarithmic and in arithmetic scales. As we already discussed, $g_{11}(\sigma_{11})$ has three terms. The first of them describes hard sphere packing effects. The second term corresponds to the depletion effect due to cluster-cluster repulsion. The third term is connected with the intramolecular correlations inside a cluster. It can be seen that at small densities the intramolecular term dominates and the contact values increases with increasing the size of clusters. Due to the depletion effect, the contact value $g_{11}(\sigma_{11})$ decreases with an increasing size of clusters at higher densities. And finally, for sufficiently dense fluids, the contact value $g_{11}(\sigma_{11})$ increases due to hard sphere packing effects. In this case, the effects from clustering are negligible.
	\begin{figure}[h]
		\centerline{
			\includegraphics [height=0.42\textwidth]{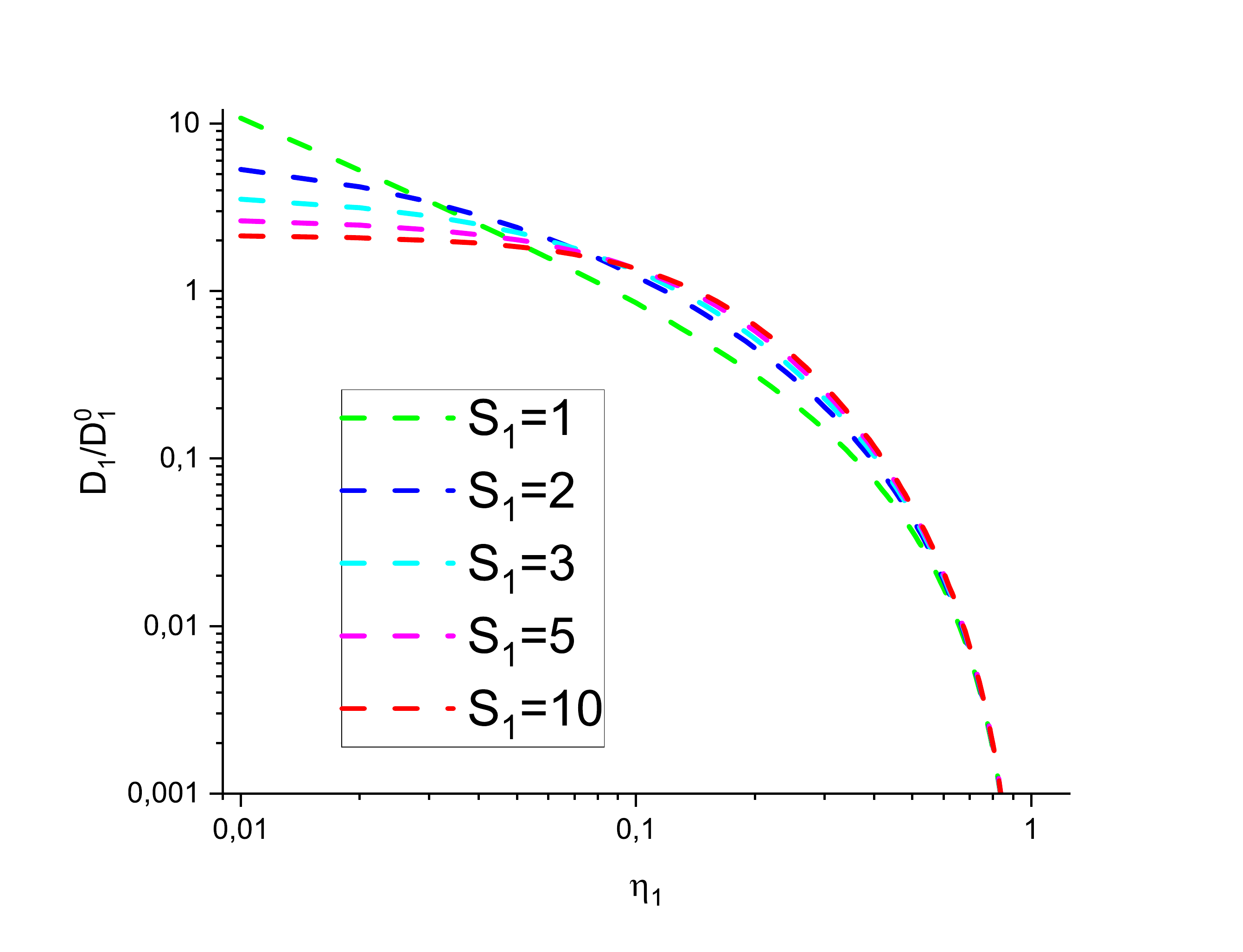}
		}
		\caption{(Colour online) The same as in figure~\ref{FIG1} but for the self-diffusion coefficient of network-like fluid.}
		\label{FIG2}
	\end{figure}
	The density dependence of self-diffusion coefficient is presented in figure~\ref{FIG2}. As we can see, this dependence is determined by the dependence of the contact value $g_{11}(\sigma_{11})$. At small densities, the self-diffusion coefficient decreases due to clustering, and afterwards at the intermediate density, the diffusion coefficient increases due to the depletion effect. For a sufficiently dense fluid it decreases due to hard sphere packing effect and does not depend on clustering. A comparison between the effects from linear chains and network-like clusters is given in figure~\ref{FIG3}. At a fixed mean cluster size, $S_1$, the effect from network-like clusters is stronger than the effect from linear chain clusters.
	\begin{figure}[h]
		\centerline{
			\includegraphics [height=0.42\textwidth]{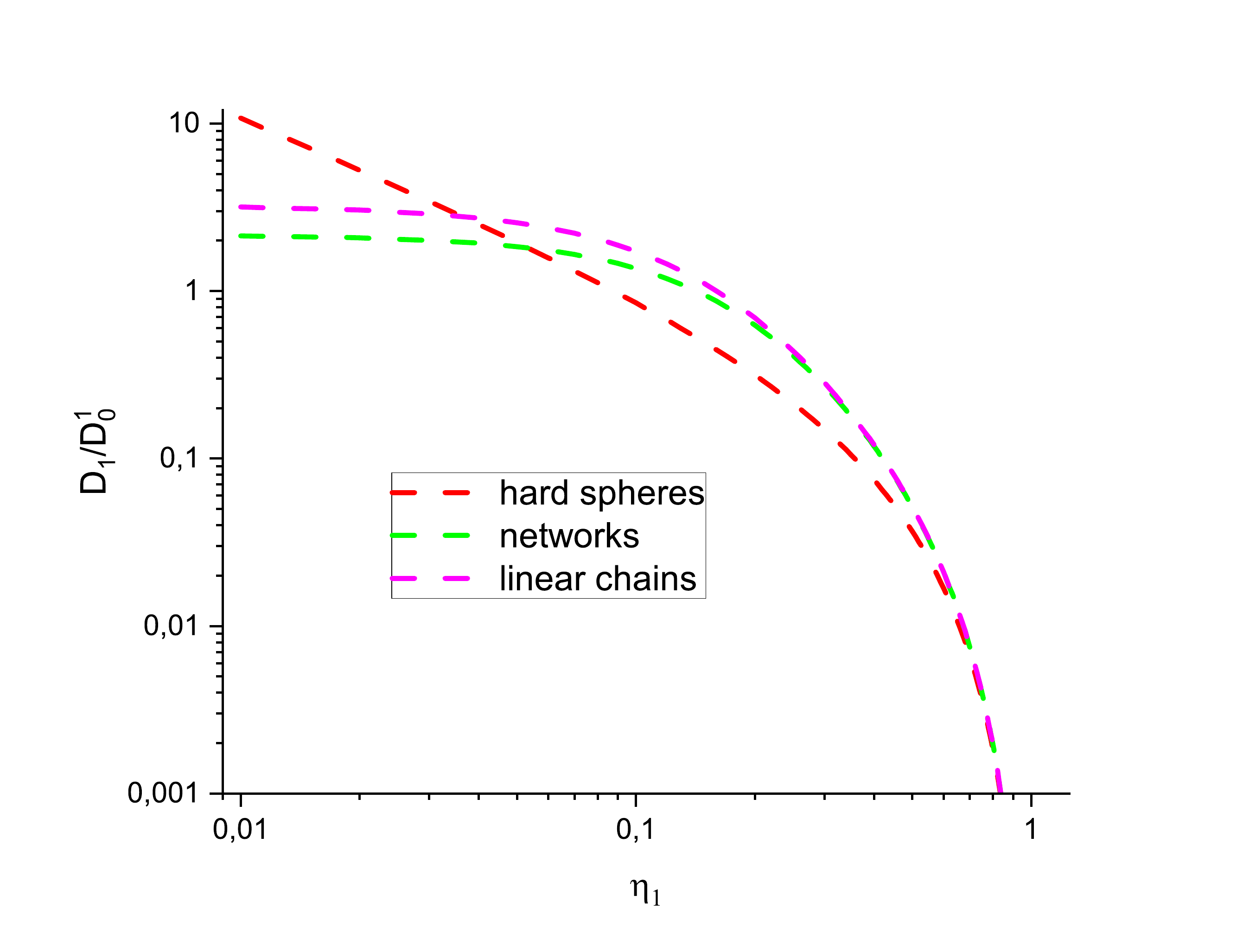}
		}
		\caption{(Colour online) The influence of type of fluid on the self-diffusion coefficient at a fixed cluster size $S_1=10$ for the bulk case.}
		\label{FIG3}
	\end{figure}

\begin{figure}[h]
	\centerline{
		\includegraphics [height=0.42\textwidth]{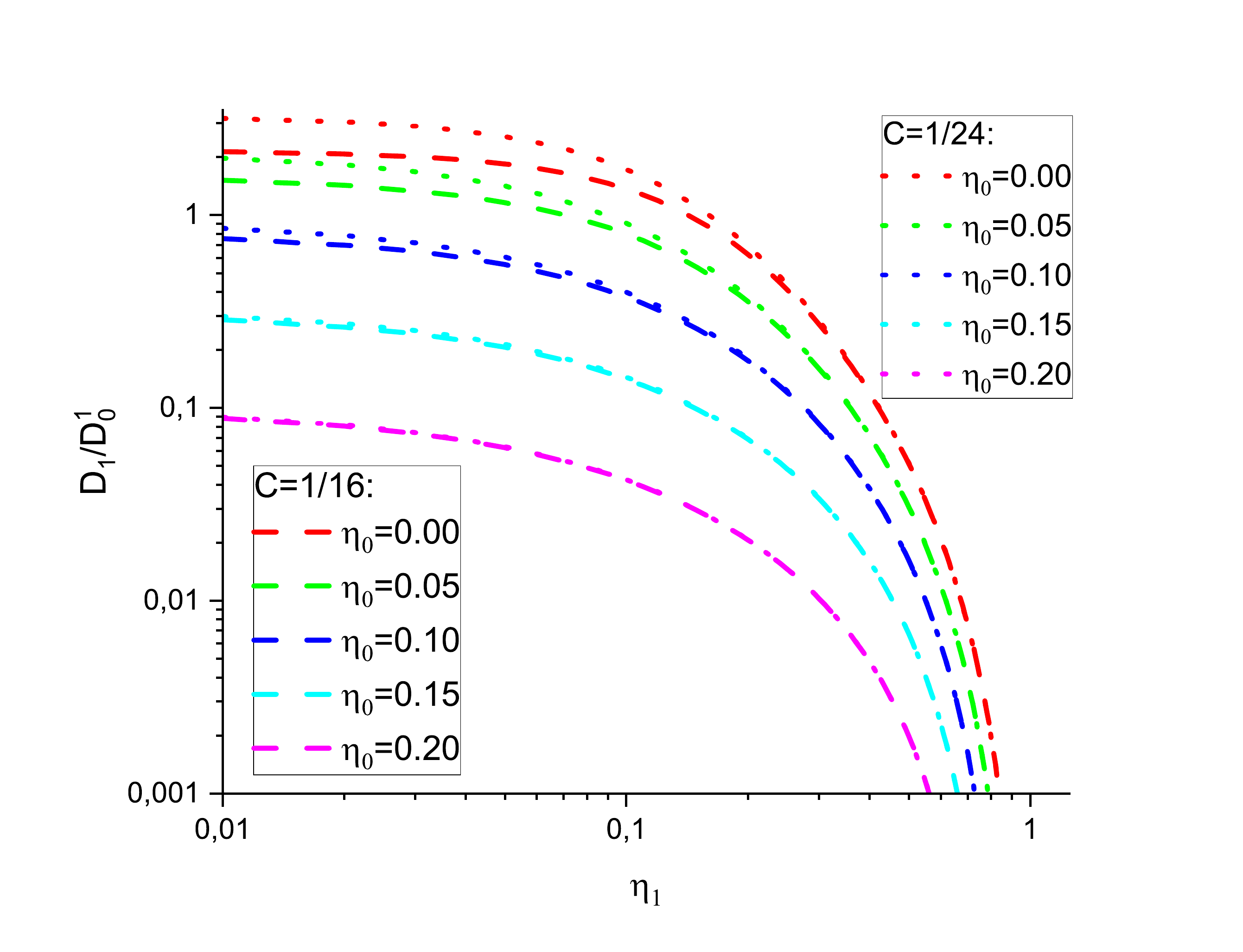}
		}
		\caption{(Colour online) The influence of matrix packing fraction $\eta_0$ and type of clustering on the self-diffusion coefficient of patchy colloids at fixed cluster size $S_1=10$ for patchy colloids in porous media. $C=1/16$ corresponds to network-like fluid, whereas $C=1/24$ corresponds to linear chain-like fluid.}
	\label{FIG4}
\end{figure}

	Now, we comment the influence of porous media. The influence of packing of porous matrix $\eta_{0}$ on the self-diffusion coefficient of a hard sphere associating fluid at a fixed cluster size, $S_1=10$, for network-like $C=1/16$ and linear chain-like clusters $C=1/24$, is presented in figure~\ref{FIG4}. The self-diffusion coefficient decreases with a decreasing porosity $\phi_{0}=1-\eta_{0}$. For a small fluid density $\eta_{1}$, the intramolecular cluster effect is modified by the influence from porous media. As we can see from~(\ref{HolKor3.1}) at $\eta_{1}\to0$,
	\begin{equation}
	{D_{n}}\to\frac{1}{\sqrt{2}}C(1-X)^{2}+\frac14{\tau}{(\tau+1)}\eta_{0}\frac{\phi_{0}}{\phi}X
	,\label{HolKor3.3}
	\end{equation}
	and the effect from the type of cluster, $C$, can be weaker than the effect of porous media. We observe from figure~\ref{FIG4} that at $\eta_{0}=0.15$, and at higher $\eta_{0}$, the effect of the type of clusters is negligible.
	
	\section{Conclusions}
	In this paper, we have generalized the Enskog theory for the description of dynamic properties of patchy colloidal fluids in disordered porous media. This generalization is based on our recent work~\cite{HolovKor20}. The patchy colloidal fluids are modelled by the system of associating hard spheres with the association due to highly directional attractive interaction that leads to the formation of clusters of colloids. Similar to reference~\cite{HolovKor20}, the extended Enskog theory includes the contact values of the fluid-fluid and fluid-matrix pair distribution functions. However, as it was noted in~\cite{HolovKor20}, the influence of porous media on the contact values of both distribution functions is described only by the geometrical porosity, $\phi_{0}$. The application of such contact values neglects the effects of trapping of fluid particles by a matrix. The concept of probe particle porosity, $\phi$, should be included into the Enskog theory for a correct description of the matrix influence. As a result, the Enskog theory was extended in~\cite{HolovKor20} by changing the contact values of the fluid-matrix and fluid-fluid pair distribution functions by their counterparts that include the dependence from geometrical porosity, $\phi_{0}$, as well as from the so-called probe particle porosity, $\phi$. Such an improvement corresponds to the SPT2b1 approximation for the description of thermodynamic properties of a hard sphere fluid in disordered porous media. It predicts the correct trends for the influence of porous media on the diffusion coefficient of a hard sphere fluid in good agreement with computer simulation data~\cite{ChangJad04}.

	In the present paper, the analogues of the contact values of the fluid-fluid and fluid-matrix pair distribution functions are introduced as an input for the extended Enskog theory for patchy colloidal fluids in disordered porous media. For this purpose, we propose the expressions~(\ref{HolKor2.12}) and~(\ref{HolKor2.13}), respectively. The first contribution in these expressions includes~(\ref{HolKor2.2}) and~(\ref{HolKor2.8}) proposed in~\cite{HolovKor20} for the description of hard sphere packing effects in disordered porous media. The second contribution in~(\ref{HolKor2.12}) and~(\ref{HolKor2.13}) is introduced as input as well. It is connected with the depletion due to the cluster-cluster and cluster-matrix repulsions and strongly depends on the cluster size $S_1$. The fluid-fluid contact value $g_{11}(\sigma_{11})$ also includes an additional term connected with the intramolecular correlations inside a cluster. This term does not depend on the presence of porous media. It is influenced by fluid packing fraction $\eta_1$, by the size of the cluster $S_1$, and by the type of clusters defined in terms of the parameter $C$. It is inversely proportional to the fluid density and is important in the so-called dilute and semi-dilute regime~\cite{VoronovLuzMinSid97}. The parameter~$C$ is different for the chain-like clusters, $C=1/24$, and for the network-like clusters, $C=1/16$. It means that the correlations in the network-like system are stronger than in the chain-like system. However, the presence of porous media may yield the influence of the type of clusters weaker in accordance to equation~(\ref{HolKor3.3}).

	We should note that the Enskog theory is fully developed only for hard sphere fluids. In our recent paper, the Enskog theory has been successfully modified  for a hard sphere fluid in random porous media. In the present paper, we generalize the Enskog theory for patchy colloidal fluids in disordered porous media. In order to check the accuracy of this description, a comparison with computer simulation results should be presented. Unfortunately, we are not aware of the simulation results for the systems of our study. Recently, molecular dynamics simulations for patchy colloidal fluids in disordered porous media have been initiated in our laboratory and we hope to compare our theoretical predictions with simulation data in future. Unfortunately, we did not find any experimental confirmation of the obtained theoretical results for the behaviour of diffusion of patchy colloids in porous media. It was shown in~\cite{VoronovLuzMinSid97} that at low density the monomers correlate mostly through cluster connectivity. This leads to a remarkable increase in surface coverage.
	At intermediate densities, the competition between the intermolecular fluid-fluid correlations and corresponding depletion effects due to fluid-fluid and fluid-matrix effects are responsible for the low-pressure behaviour and possible separation of chain-like molecules in host matrices~\cite{VakBadi06}. Probably, similar mechanism of dynamical behaviour of patchy colloids in porous media could occur.

	Finally, we should note that in the present paper we consider the mean cluster size $S_1$ as an external parameter. Hence, we do not discuss the clustering process itself and assume that the cluster composition remains fixed. In a more general case, $S_1$, should be considered as a function of fluid density, $\eta_1$, temperature $T$, and the parameters of porous media such as, $\eta_0$ and $\tau$. We plan to consider such a setup in a separate work.
	
	\section*{Acknowledgement}
	M. H. acknowledges support from the National Research Foundation of Ukraine (Grant agreement No~50/02.2020). We thank Orest Pizio for careful reading of the manuscript and very useful comments.

	\newpage
	
	\ukrainianpart
	
	\title{Вплив ефектів кластиризації на дифузiю плямистих колоїдів у невпорядкованому пористому середовищi}%
	\author{М. Головко, М. Корвацька }
	\address{
		Інститут фізики конденсованих систем Національної академії наук України, вул. Свєнціцького, 1, \\79011 Львів, Україна
	}

	\makeukrtitle
	
	\begin{abstract}
		Теорія Енскога узагальнена для опису коефіцієнта самодифузії плинів плямистих колоїдів у невпорядкованих пористих середовищах. Ця теорія включає контактні значення плин-плин та плин-матриця для парних функцій розподілу, які були модифіковані, щоби включити залежність від пористості $\phi$ для правильного опису ефектів захоплення частинок плину матрицею. Запропоновані вирази для модифікованих контактних значень плин-плин і плин-матриця парних функцій розподілу включають три вклади, зокрема вклад твердих сфер, отриманий нами в попередній роботі [Holovko M.~F., Korvatska M.~Ya., Condens. Matter Phys., 2020, \textbf{23}, 23605], внесок від ефекту збіднення, пов'язаний з відштовхуванням кластер-кластер та кластер-матриця та вклад, повязаний з внутрішньомолекулярною кореляцією всередині кластера. Показано, що останній вклад приводить до значного зменшення коефіцієнта самодифузії при низьких густинах плину. Із зменшенням пористості цей ефект ослаблюється. Для проміжних густин ефект збіднення частинок призводить до збільшення коефіцієнта самодифузії. Для досить густих плинів за рахунок ефекту твердих сфер коефіцієнт самодифузії сильно знижується. Досліджено та обговорено вплив розмірів та типів кластеризації, а також вплив параметрів пористого середовища.

		\keywords
		плямисті колоїди, невпорядковані пористі середовища, теорія Енскога, коефіцієнт самодифузії, геометрична пористість, пористість пробної частинки
	\end{abstract}
\end{document}